\documentclass[aps,prl,superscriptaddress,twocolumn]{revtex4-2}
\usepackage{amssymb,amsmath,amsfonts}
\usepackage{graphicx}
\usepackage{theorem}
\usepackage{dcolumn}
\usepackage{braket}
\usepackage{bm}
\usepackage{comment}
\usepackage{setspace}
\usepackage{eurosym}
\usepackage{amsfonts}
\usepackage{amsmath}
\usepackage{mathrsfs}
\usepackage{amssymb}
\usepackage{graphicx}
\usepackage{float}
\usepackage{comment}
\usepackage{bm}
\usepackage{color}
\usepackage[colorlinks=true,linkcolor=blue,anchorcolor=blue,citecolor=blue,urlcolor=blue]{hyperref}
\usepackage[normalem]{ulem}
\usepackage{siunitx}
\usepackage[tracking]{microtype}
\usepackage[version=4]{mhchem}
\usepackage{elements}

\begin{document}

\title{Dynamical Transition of Quantum Vortex-Pair Annihilation in a
Bose-Einstein Condensate}
\author{Toshiaki Kanai}
\affiliation{Department of Physics, Washington University, St.\,Louis, Missouri 63130, USA}
\author{Chuanwei Zhang}
\email{Email: chuanwei.zhang@wustl.edu}
\affiliation{Department of Physics, Washington University, St.\,Louis, Missouri 63130, USA}

\begin{abstract}
Understanding the elementary mechanism for the dissipation of vortex energy
in quantum liquids is one central issue in quantum hydrodynamics, such as
quantum turbulence in systems ranging from neutron stars to atomic
condensates. In a two-dimensional (2D) Bose-Einstein condensate (BEC) at
zero temperature, besides the vortex drift-out process from the boundary,
vortex-antivortex pair can annihilate in the bulk, but controversy remains
on the number of vortices involved in
the annihilation process. We find there exists a dynamical transition from
four-body to three-body vortex annihilation processes with the time
evolution in a boundary-less uniform quasi-2D BEC. Such dynamical transition
depends on the initial vortex pair density, and occurs when the sound waves
generated in the vortex annihilation process surpass a critical energy. With
the confinement along the third direction is relaxed in a quasi-2D BEC, the critical sound wave energy
decreases due to the 3D vortex line curve and reconnection, shifting the
dynamical transition to the early time. Our work reveals an elementary
mechanism for the dissipation of vortex energy that may help understand
exotic matter and dynamics in quantum liquids.
\end{abstract}

\maketitle

Quantum vortices, ubiquitous topological excitations, play an important role
in diverse quantum phenomena such as Berezinskii-Kosterlitz-Thouless
transition~\cite{stock_observation_2005,kruger_critical_2007,
schweikhard_vortex_2007, hadzibabic_berezinskiikosterlitzthouless_2006,
nazarenko_bose-einstein_2014, sunami_observation_2022}, Kibble--Zurek
mechanism~\cite{scherer_vortex_2007, goo_defect_2021,
rysti_suppressing_2021, ko_kibblezurek_2019,weiler_spontaneous_2008,
kanai_flows_2018, kanai_merging_2019, kanai_torque_2020}, and quantum turbulence~\cite%
{barenghi_quantum_2024}. In quantum liquids, the quantization of the
superflow circulation around the vortex makes the vorticity robust against
decay. The understanding of vortex motion and their decaying mechanism is
thus fundamentally important and has attracted much attention recently,
particularly in 2D quantum turbulence~\cite{lucas_sound-induced_2014, 
johnstone_evolution_2019, gauthier_giant_2019, sachkou_coherent_2019, 
simula_emergence_2014, billam_onsager-kraichnan_2014, groszek_onsager_2016, 
seo_observation_2017, easton_vortex_2023, reeves_turbulent_2022, groszek_decaying_2020,
karl_strongly_2017, cidrim_controlled_2016, groszek_vortex_2018, frishman_turbulence_2018,
yu_theory_2016, schole_critical_2012}.
In a 2D BEC,  quantum vortices may disappear via two processes: vortex drifting-out from 
the boundary and vortex-pair annihilation in the bulk~\cite{kwon_relaxation_2014}. 
Here the vortex pair represents a pair of vortex and antivortex with positive and
negative single quantum of circulations.

\begin{figure}[hb!]
\centerline{\includegraphics[width=3.0in]{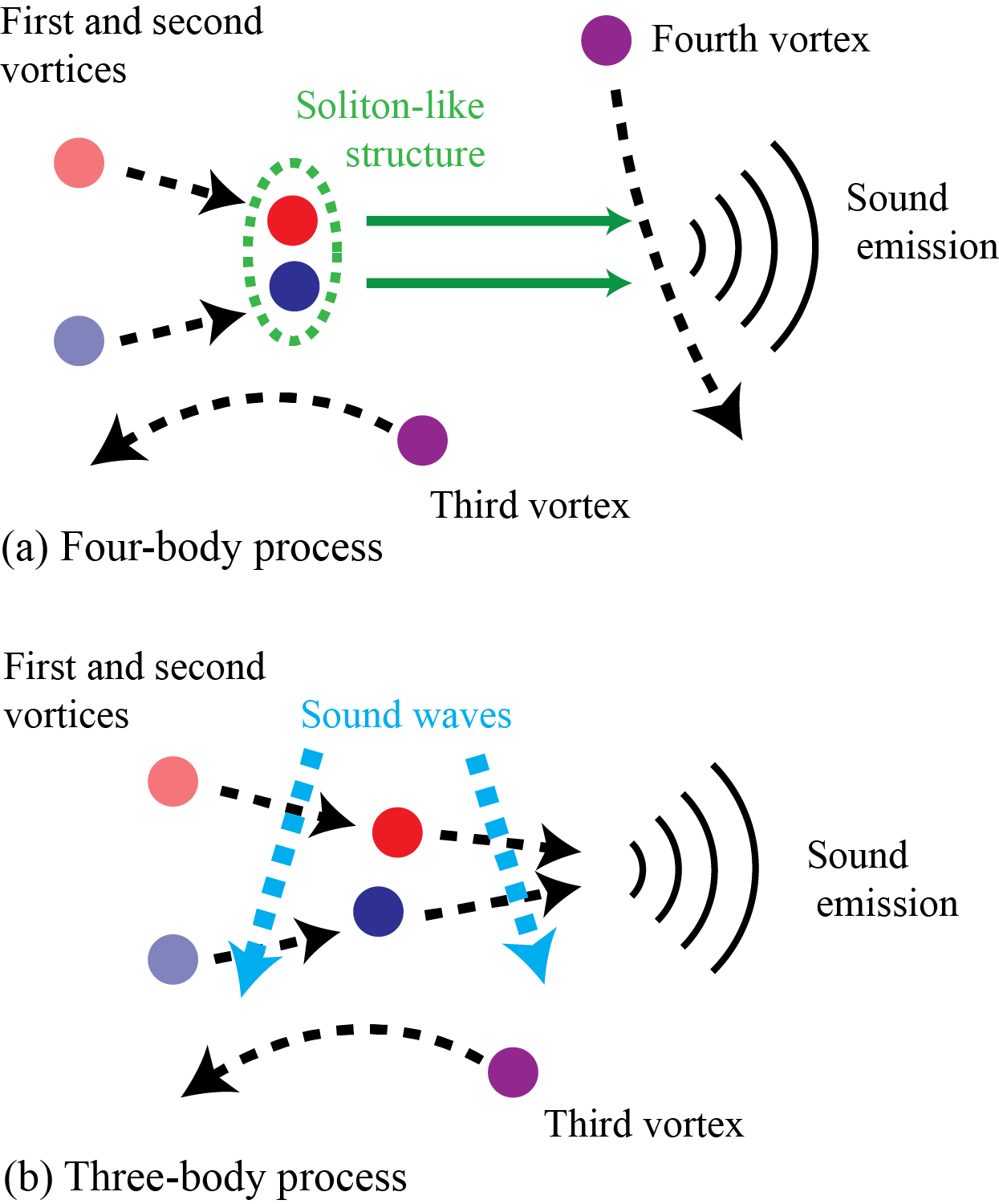}}
\caption{Schematics of the vortex-pair annihilation process. (a) Four-body
process: the first and second vortices form a soliton-like structure. The
third vortex is needed to conserve the vortex energy and momentum. This
soliton-like structure is robust in uniform 2D BECs, and the fourth vortex
is required to break the structure into sound waves. (b) Three-body process:
the pair of the first and second vortices dissipates to sound waves via the
interaction with the third vortex and sound waves.}
\label{fig:vortex-pair-annihilation}
\end{figure}

So far the basic mechanism for the vortex-pair annihilation remains
inconclusive. While vortex-pair annihilation was initially considered as a
two-body process \cite{kwon_relaxation_2014, stagg_generation_2015}, later
studies utilizing the purely 2D Gross-Pitaevskii equation (GPE) showed that
it is a four-body process at zero temperature~\cite{groszek_onsager_2016,
baggaley_decay_2018, cidrim_controlled_2016}, where the vortex pair forms a
soliton-like structure~\cite{smirnov_dynamics_2012, smirnov_scattering_2015,
nazarenko_freely_2007, rickinson_diffusion_2018} (i.e., a pair with a short vortex distance) through a
three-body interaction with the third vortex. This soliton-like structure is
robust, and the fourth vortex is needed to dissipate it to sound waves, as
illustrated in Fig.~\ref{fig:vortex-pair-annihilation}(a). However, recent
work utilizing 3D GPE demonstrated that the vortex-pair annihilation is a
three-body process (Fig.~\ref{fig:vortex-pair-annihilation}(b)) in a
uniform quasi-2D BEC \cite{kanai_true_2021} with a strong
confinement along the third direction. These different results raise the
natural question regarding the fundamental mechanism behind the vortex-pair
annihilation, particularly in the dimensional crossover from 2D to 3D for a
realistic quasi-2D BEC.

\begin{figure*}[th]
\centering
\includegraphics[width=0.9\textwidth]{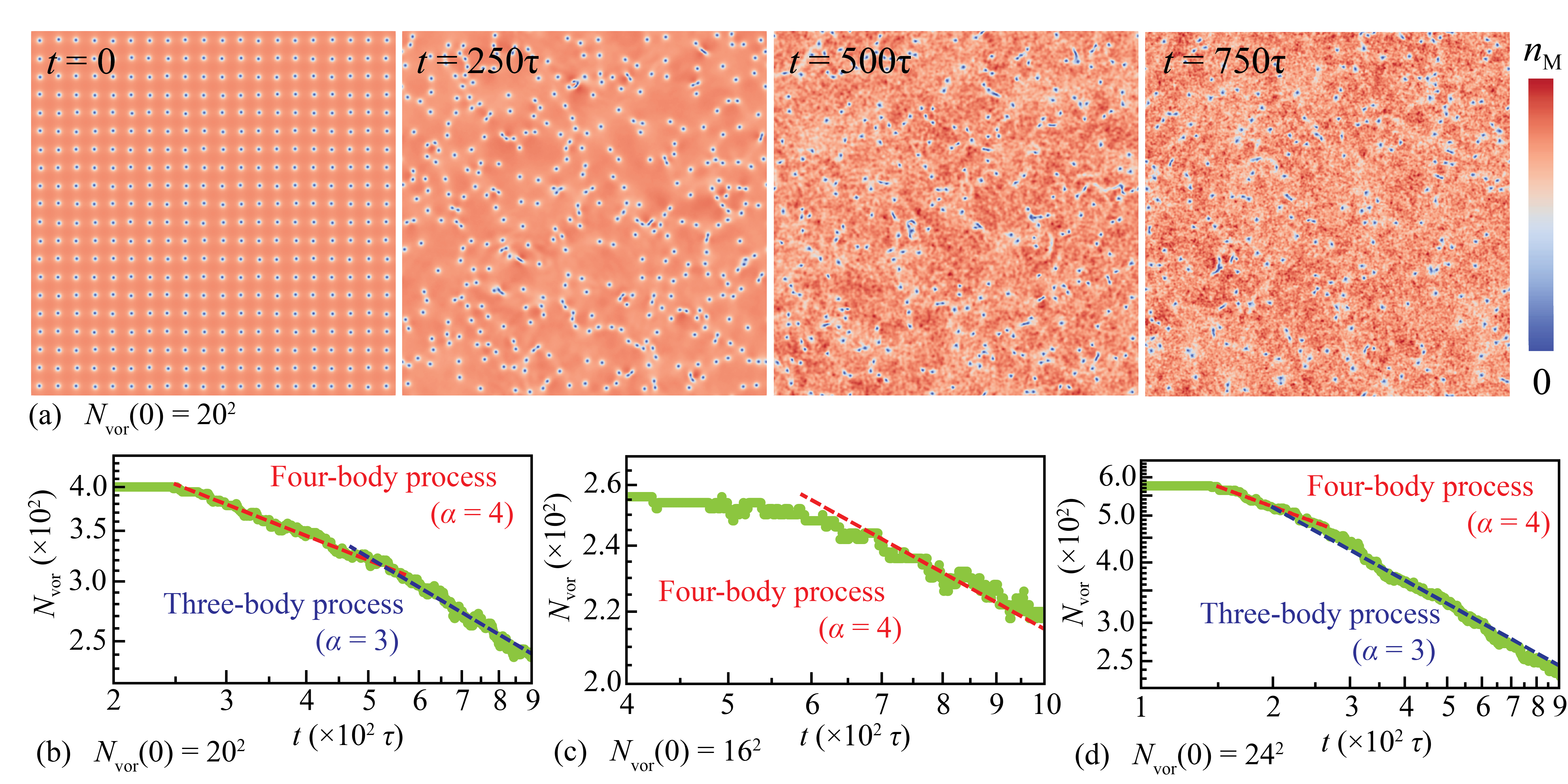}
\caption{(a) Snapshots of density profile in a representative case with $%
L_{z}=2\protect\xi $ and the initial vortex number $N_{\mathrm{vor}%
}(0)=20^{2}$. These plots are on the cross-section at $z=0$. The maximum
value of the color bar is $n_{\mathrm{M}}=2\times 10^{-5}N/\protect\xi ^{3}$%
. (b-d) Time evolution of the vortex number decay. The initial vortex number
is  $N_{\mathrm{vor}}(0)=20^{2}$ (b), $N_{\mathrm{vor}}(0)=16^{2}$ (c), and $%
N_{\mathrm{vor}}(0)=24^{2}$ (d). The red and blue dashed lines correspond to
four-body and three-body processes, respectively.}
\label{fig:density_evolution}
\end{figure*}

In this Letter, we address this important question by simulating vortex
decay dynamics in a boundary-less uniform quasi-2D BEC utilizing the 3D GPE
with different confinements along the third direction. Here the
boundary-less condition prohibits the vortex drifting-out process in the 2D
plane. Our main results are:

\textit{i}) The vortex annihilation process is governed by the sound wave intensity
characterized by a critical energy $E_{c}$, below (above) which the
annihilation process is four-body (three-body). With a low initial vortex
density, the vortex-pair annihilation remains a four-body process during the
dynamical evolution due to the low sound wave energy generated in the
annihilation process.

\textit{ii}) In the high initial vortex density region, the vortex annihilation
starts from the four-body process. However, the generated sound wave energy
from the vortex annihilation exceeds $E_{c}$ after certain time $t$, leading to
a dynamical transition \cite{ma_phase_2019} from four-body to three-body processes.

\textit{iii}) The critical sound wave energy decreases for weakening confinement
 along the third direction. As the confinement 
becomes weaker, the vortex lines can curve and reconnect along the third dimension, which
reduce the effective critical sound wave energy for annihilating the vortex
pairs, shifting the dynamical transition to an early time.

\paragraph*{Description of a quasi-2D BEC and numerical method.}

\begin{figure}[t]
\centering
\includegraphics[width=0.43\textwidth]{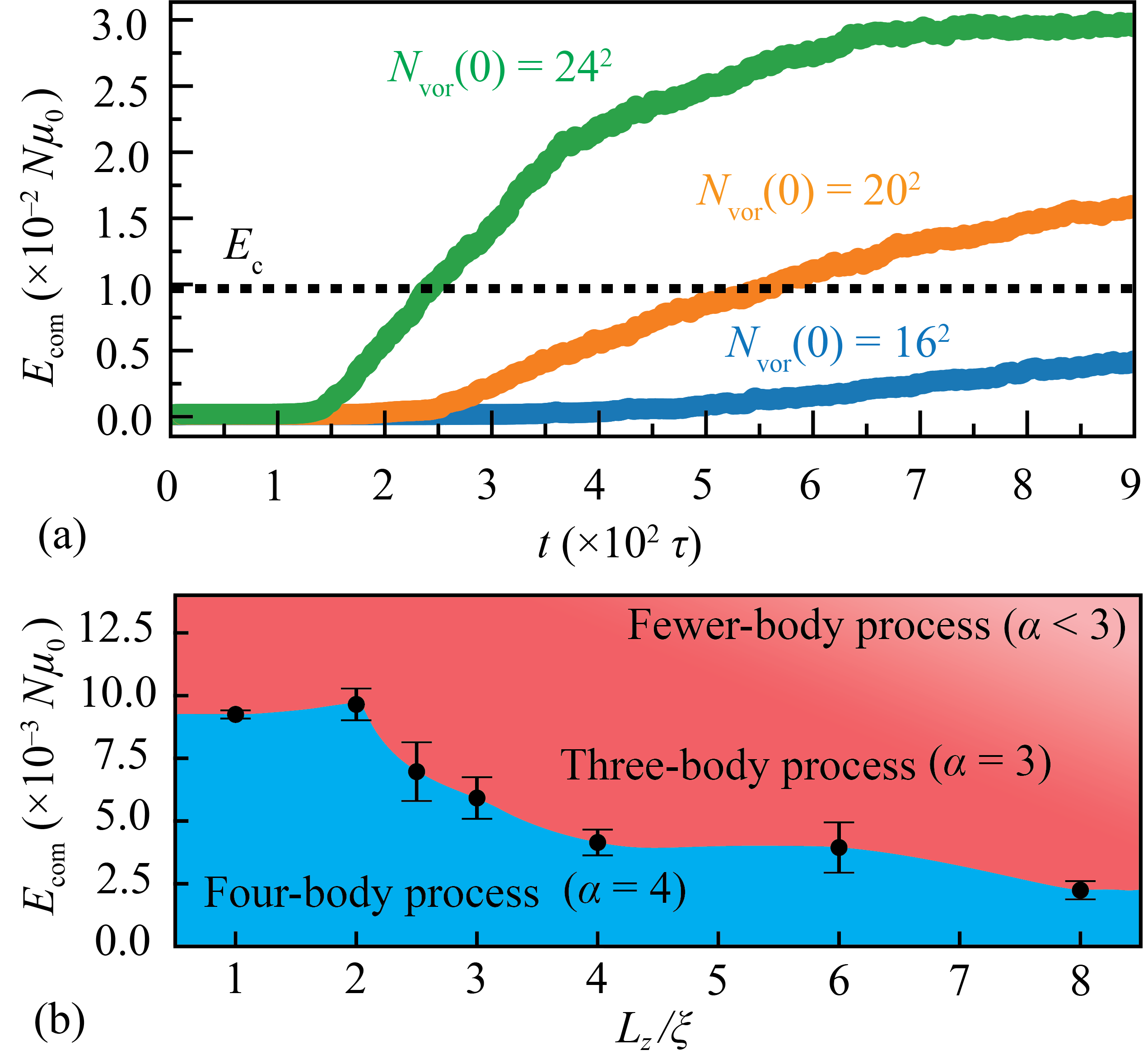}
\caption{ (a) Time evolution of the compressible kinetic energy $E_{\mathrm{%
com}}$ with various initial vortex numbers $N_{\mathrm{vor}}(0)$ under the same $L_{z}=2\protect\xi $. The dashed line
represents the critical energy $E_{\mathrm{C}}$ determined from $N_{\mathrm{vor}}(0)=20^{2}$. The three-body scaling is observed
when $E_{\mathrm{com}}>E_{\mathrm{C}}$. (b) Dependence of $E_{\mathrm{C}}$ on the $z$-axis confinement $L_{z}$. $N_{\mathrm{vor}%
}(0)=20^{2}$. The boundary between three and four-body processes is drawn for illustration by connecting $E_{\mathrm{C}}$. 
With larger $E_{\mathrm{com}}$, the vortex-pair
annihilation may involve the two-body process, where $\protect%
\alpha $ is smaller than $3$. }
\label{fig:ene_com}
\end{figure}

We simulate the dynamics of a quasi-2D BEC by the 3D GPE~\cite%
{pethick_bose-einstein_2008}: 
\begin{equation}
i\hbar \frac{\partial \Psi (\bm{r},t)}{\partial t}=-\frac{\hbar ^{2}}{2m}%
\nabla ^{2}\Psi +U(z)\Psi +C|\Psi |^{2}\Psi -\mu \Psi ,
\end{equation}%
where $\Psi $ is the macroscopic wave function, $\hbar $ is the reduced
Plank constant, $m$ is the atom mass, and $C=4\pi \hbar ^{2}a_{s}/m$ is the
interaction strength with the $s$-wave scattering length $a_{s}$. For simplicity, we assume the
quasi-2D BEC is confined in the $z$-direction by a box potential 
\begin{equation}
U(z)=\left\{ 
\begin{array}{ll}
\infty  & \hspace{10pt}\left\vert z\right\vert \geq L_{z}/2 \\ 
0 & \hspace{10pt}\left\vert z\right\vert <L_{z}/2.%
\end{array}%
\right. 
\end{equation}%
The periodic boundary condition is assumed in the $x$-$y$ plane so that
quantum vortices may disappear only through the vortex-pair annihilation
process. Kelvin waves, \textit{i.e.}, transverse excitations along a vortex
line, are suppressed in thin BECs with a small $L_{z}$ \cite%
{rooney_suppression_2011}, and such suppression is often considered as the
criteria of 2D in the context of quantum turbulence~\cite%
{chomaz_emergence_2015}.

The BEC density $n$ is given by $n=\left\vert \Psi \right\vert ^{2}$, and
the velocity field $\bm{v}$ is computed by $\bm{v}(\bm{r},t)=\frac{\hbar }{%
2mi}\frac{\Psi ^{\ast }\bm{\nabla}\Psi -\Psi \bm{\nabla}\Psi ^{\ast }}{n}$.
The BEC dynamics has the healing length scale $\xi ={\hbar }/\sqrt{2m\mu _{0}%
}$ and time scale $\tau ={\hbar }/{\mu _{0}}$ with the equilibrium chemical
potential $\mu _{0}=Cn_{0}$ and the average BEC density $n_{0}$. We solve
the GPE by the fourth-order Runge-Kutta-Gill's method~\cite%
{abramowitz_handbook_2013} with the spatial step $\Delta x=0.25\xi $ and the
time step $\Delta t=0.1\times 0.25^{2}\tau $. The computational domain size
is $L_{x}\times L_{y}\times L_{z}=\left[ 0,256\xi \right] \times \left[
0,256\xi \right] \times \left[ -L_{z}/2,L_{z}/2\right] $. In our
simulations, $L_{z}$ varies from $\xi $ to $8\xi $, which are thin enough
for the 2D criteria. For the initial state, we arrange $N_{\mathrm{vor}%
}(0)/2$ positive and $N_{\mathrm{vor}}(0)/2$ negative vortices alternatively as
a square lattice in the $x$-$y$ plane, yielding the vortex
density $N_{\mathrm{vor}}(0)/\left( 256\xi \right) ^{2}$. To reduce the
initial density fluctuations, we generate the initial state by the imaginary
time evolution with the pinning potential located at vortex locations, hence
the initial state has no sound waves~\cite{banerjee_structural_2022}.

\paragraph*{Dynamical transition from four-body to three-body processes.}

Fig.~\ref{fig:density_evolution} (a) shows the density evolution of a BEC
with $L_{z}=2\xi $ and the initial vortex number $N_{\mathrm{%
vor}}(0)=20^{2}$. At $t=0$, the BEC has no sound waves, and
the vortices are placed as a square lattice with the lattice constant $\sqrt{L_{x}L_{y}/N_{%
\mathrm{vor}}}=256\xi /20$. Here we slightly shift the vortices randomly to
break the discrete translational symmetry. As the BEC evolves, the vortex
pairs annihilate, emitting sound waves. The sound waves are less in the
early time $t=250\tau $, but the BEC becomes strongly wavy in a later time
 $t=750\tau $.

The number of vortices involved in the vortex decay process can be
characterized by the scaling of the vortex number decay~\cite%
{baggaley_decay_2018, groszek_onsager_2016}. For a $\alpha $-body vortex
decay process, the rate equation of the vortex number $N_{\mathrm{vor}}$ is
given by 
\begin{equation}
\frac{dN_{\mathrm{vor}}(t)}{dt}\propto -N_{\mathrm{vor}}^{\alpha },
\end{equation}%
therefore the vortex number scales as 
\begin{equation}
N_{\mathrm{vor}}(t)\propto t^{-{1}/\left( \alpha -1\right) }.
\end{equation}%
Fig.~\ref{fig:density_evolution}(b) shows the vortex number decay
corresponding to Fig.~\ref{fig:density_evolution}(a). After vortex pairs
start to annihilate, the vortex number follows the scaling $N_{\mathrm{vor}%
}(t)\propto t^{-1/3}$ at the early time, corresponding to the four-body
process. After certain time of evolution, the vortex number scaling changes
to $N_{\mathrm{vor}}\propto t^{-1/2}$, corresponding to the three-body
process. The transition depends on the initial vortex number (i.e.,
vortex density): there is only four-body process for $N_{\mathrm{vor}}(0)=16^{2}$
(Fig. \ref{fig:density_evolution}(c)), while the transition occurs earlier
for $N_{\mathrm{vor}}(0)=24^{2}$ (Fig. \ref{fig:density_evolution}(d)). Such
dynamical transition from four-body to three-body processes have not been
discussed before and understanding the mechanism behind it is
the main focus hereafter.

\paragraph{Dynamical transition induced by increasing sound waves.}

At finite temperatures, the thermal excitations dampening the kinetic
energy may change the scaling of $N_{\mathrm{vor}}$~\cite%
{baggaley_decay_2018, groszek_onsager_2016}, but the GPE model here has no
thermal excitations. Sound waves can extract energy from quantum vortices~%
\cite{nazarenko_freely_2007}, and a complementary test of vortex-pair
annihilation shown later verifies the damping effect. In our case, intense
sound waves generated by the pair-annihilation in the early stage dampen the
vortex energy later, varying the scaling to a three-body process illustrated
in Fig.~\ref{fig:vortex-pair-annihilation}(b).

\begin{figure}[t]
\centering
\includegraphics[width=0.48\textwidth]{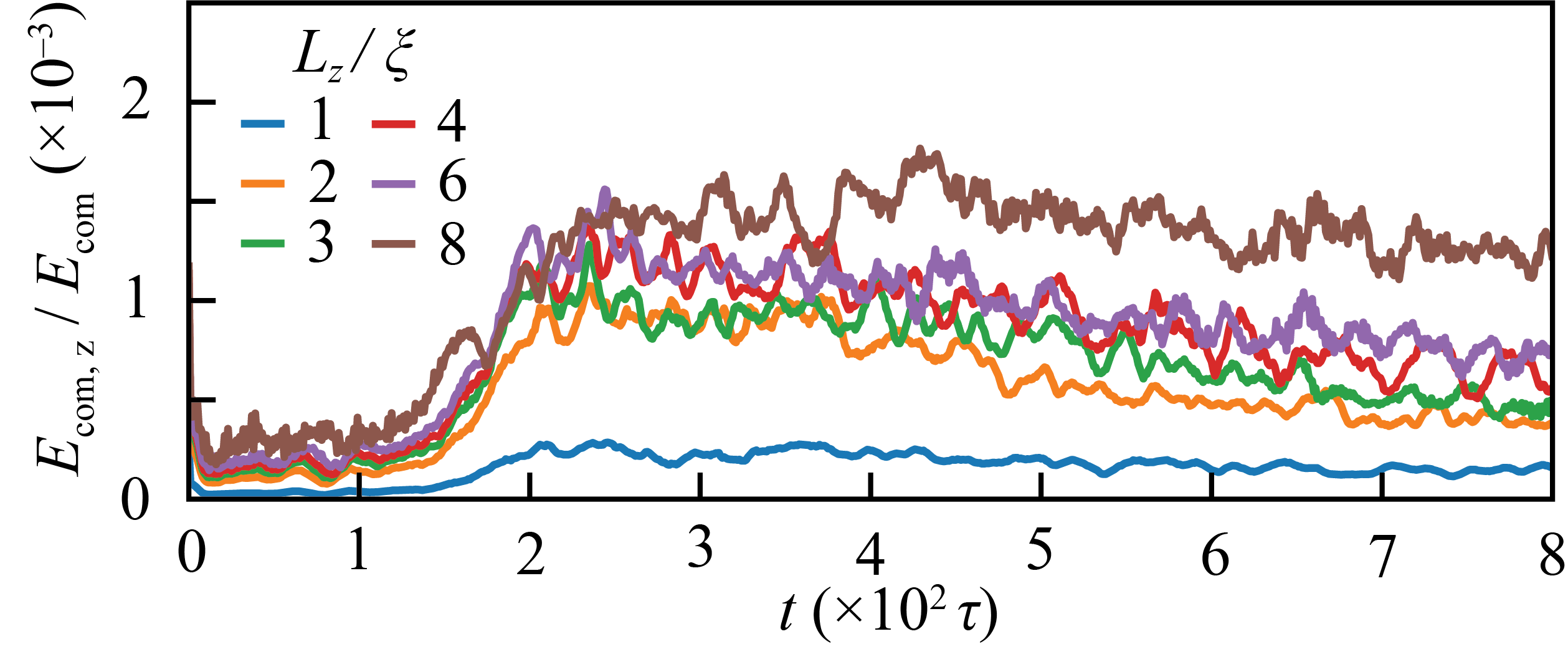}
\caption{ Time evolution of the fraction of the $z$ component contribution
$E_{\mathrm{com}, z} / E_{\mathrm{%
com}}$ with various confinement $L_z$. }
\label{fig:transition}
\end{figure}

To characterize the sound wave intensity, we decompose the velocity field as 
$\bm{v}=\bm{v}_{\mathrm{com}}+\bm{v}_{\mathrm{inc}}$ with the incompressible
velocity $\bm{v}_{\mathrm{inc}}$ and the compressible velocity $\bm{v}_{%
\mathrm{com}}$, satisfying $\bm{\nabla}\cdot \left( \sqrt{n}\bm{v}_{\mathrm{%
inc}}\right) =0$ and $\bm{\nabla}\times \left( \sqrt{n}\bm{v}_{\mathrm{com}%
}\right) =0$, respectively. Then the compressible kinetic energy $E_{\mathrm{%
com}}$ and the incompressible kinetic energy $E_{\mathrm{inc}}$ can be
computed by~\cite{horng_two-dimensional_2009, numasato_direct_2010, bradley_energy_2012} 
\begin{eqnarray}
E_{\mathrm{com}} &=&\frac{m}{2}\int n\left\vert \bm{v}_{\mathrm{com}%
}\right\vert ^{2}dV \\
E_{\mathrm{inc}} &=&\frac{m}{2}\int n\left\vert \bm{v}_{\mathrm{inc}%
}\right\vert ^{2}dV.
\end{eqnarray}%
$E_{\mathrm{com}}$ and $E_{\mathrm{inc}}$ are associated with sound waves
and vortices, respectively~\cite{horng_two-dimensional_2009,
bradley_energy_2012}. Fig.~\ref{fig:ene_com} (a) shows the $E_{\mathrm{com}}$
evolution with various initial vortex numbers. $%
E_{\mathrm{com}}$ steadily increases as more vortex-pairs annihilate. In the
case of $N_{\mathrm{vor}}(0)=20^{2}$, the three-body process emerges after
about $t=520\tau $, when the compressible kinetic energy is about $E_{%
\mathrm{com}}\approx 9.7\times 10^{-3}N\mu _{0}$ with the total particle
number $N=\int ndV$. This compressible kinetic energy is defined as the
critical energy $E_{\mathrm{C}}$. With $N_{\mathrm{vor}}(0)=16^{2}$, $E_{%
\mathrm{com}}$ is always lower than $E_{\mathrm{C}}$ in our simulation, and
the scaling of $N_{\mathrm{vor}}$ follows the four-body process during the
evolution (Fig.~\ref{fig:density_evolution} (c)). For $%
N_{\mathrm{vor}}(0)=24^{2}$, $E_{\mathrm{com}}$ quickly increases and
becomes higher than $E_{\mathrm{C}}$, inducing the dynamical transition
early in time (Fig.~\ref{fig:density_evolution} (d)). $E_{\mathrm{C}}$ is estimated
from $N_{\mathrm{vor}}(0)=20^{2}$ but matches the
transition time for $N_{\mathrm{vor}}(0)=24^{2}$, indicating
that the critical energy is mostly independent of the initial vortex number
in our parameter range.

\paragraph{Confinement dependence in the quasi-2D BEC.}

When $L_{z}$ increases, we expect a crossover from 2D to quasi-2D and then 3D for the
BEC. For each $L_{z}$, we sample three cases from different initial states
with the same initial vortex number $N_{\mathrm{vor}}(0)=20^{2}$ and compute
the arithmetic-averaged critical energy $\bar{E}_{\mathrm{C}}$ shown in Fig.~%
\ref{fig:ene_com} (b). We see $\bar{E}_{\mathrm{C}}$ is almost a constant
below $L_{z}=2\xi $ and starts decreasing beyond $L_{z}=2\xi $. The decrease
of $\bar{E}_{\mathrm{C}}$ cannot attribute to Kelvin waves since they are
still suppressed even for $L_{z}=8\xi $. One intuitive guess is the
increase of the sound waves in the $z$-direction, which may be evaluated by
the compressible kinetic energy in the $z$-direction 
$
E_{\mathrm{com},z}=\frac{m}{2}\int n\left\vert \bm{v}_{\mathrm{com}}\cdot %
\bm{\hat{e}}_{z}\right\vert ^{2}dV
$
with the unit vector in the $z$-direction $\hat{e}_{z}$. Fig.~\ref{fig:ene_com}
(b) shows the fraction $E_{\mathrm{com},z}/E_{\mathrm{com}}$. In our parameter range $\xi \leq L_{z}\leq 8\xi $, $E_{\mathrm{%
com},z}$ is always less than $0.2\%$ of $E_{\mathrm{com}}$, demonstrating
that the $z$-component of sound waves is negligible. Therefore, the change of
the critical energy should come from the vortex contribution.

\begin{figure}[t]
\centering
\includegraphics[width=0.43\textwidth]{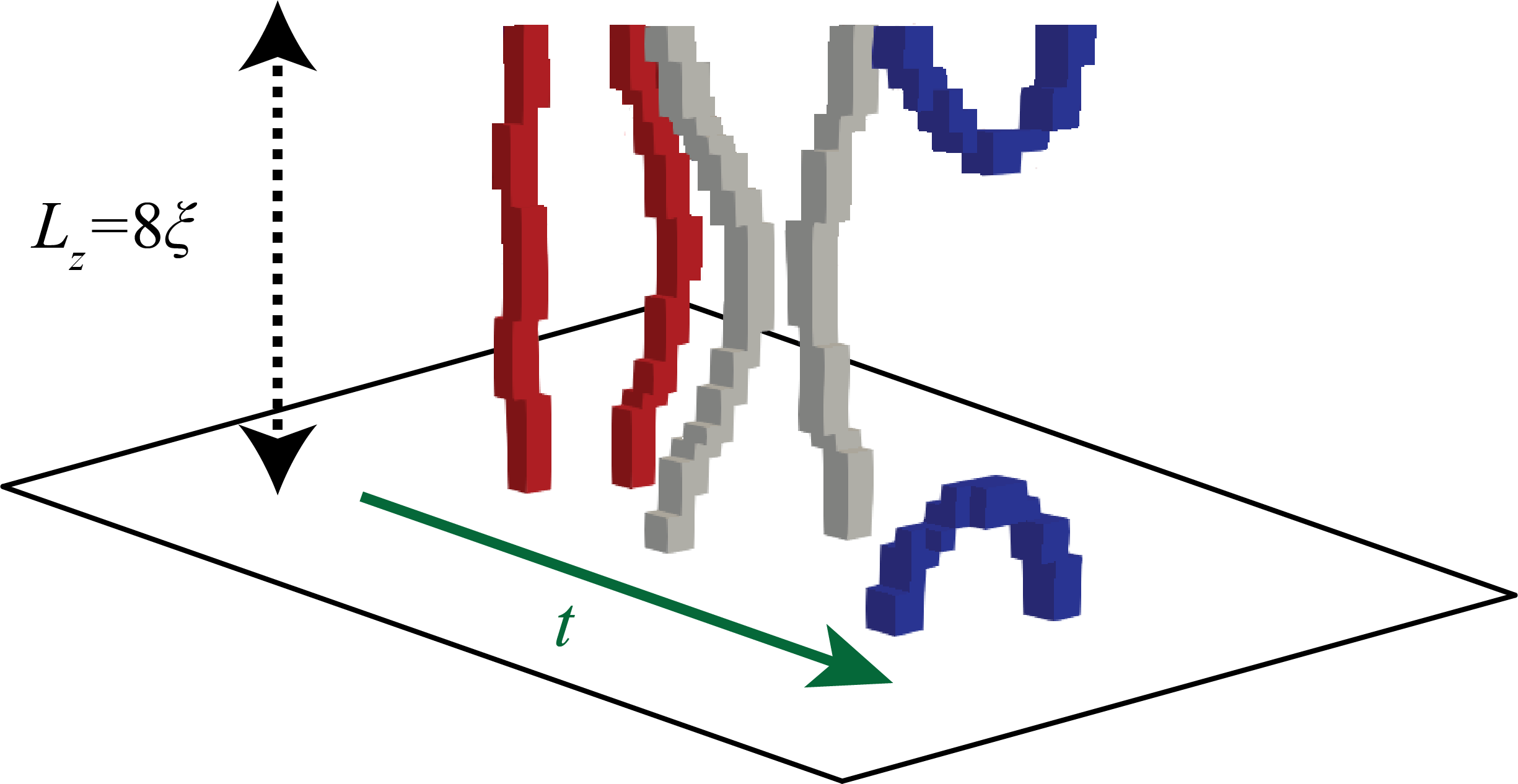}
\caption{ Vorticity evolution of a vortex reconnection event in a BEC
simulation. $L_{z}=8\protect\xi $. The red, white, and blue
lines represent the vorticity at $t=350\protect\tau $, $355\protect\tau $,
and $360\protect\tau $, respectively. }
\label{fig:reconnection}
\end{figure}

We find that the decrease of $\bar{E}_{\mathrm{C}}$ with increasing $L_{z}$ is due to the emergence
of vortex reconnection channel. As $L_{z}$
increases, vortices can be easily curved, and the vortex distance may vary
in the $z$-direction. Thus the soliton-like structure becomes unstable,
enhancing the vortex reconnection channel that makes the sound wave easier
to annihilate the vortex pair. Fig.~\ref{fig:reconnection} shows snapshots
of a vortex reconnection event in a BEC with $L_{z}=8\xi $. To
quantitatively characterize this effect, we analyze the incompressible
kinetic energy of the $z$-component 
\begin{equation}
E_{\mathrm{inc},z}=\frac{m}{2}\int n\left\vert \bm{v}_{\mathrm{inc}}\cdot %
\bm{\hat{e}}_{z}\right\vert ^{2}dV.
\end{equation}

\begin{figure}[t]
\centering
\includegraphics[width=0.43\textwidth]{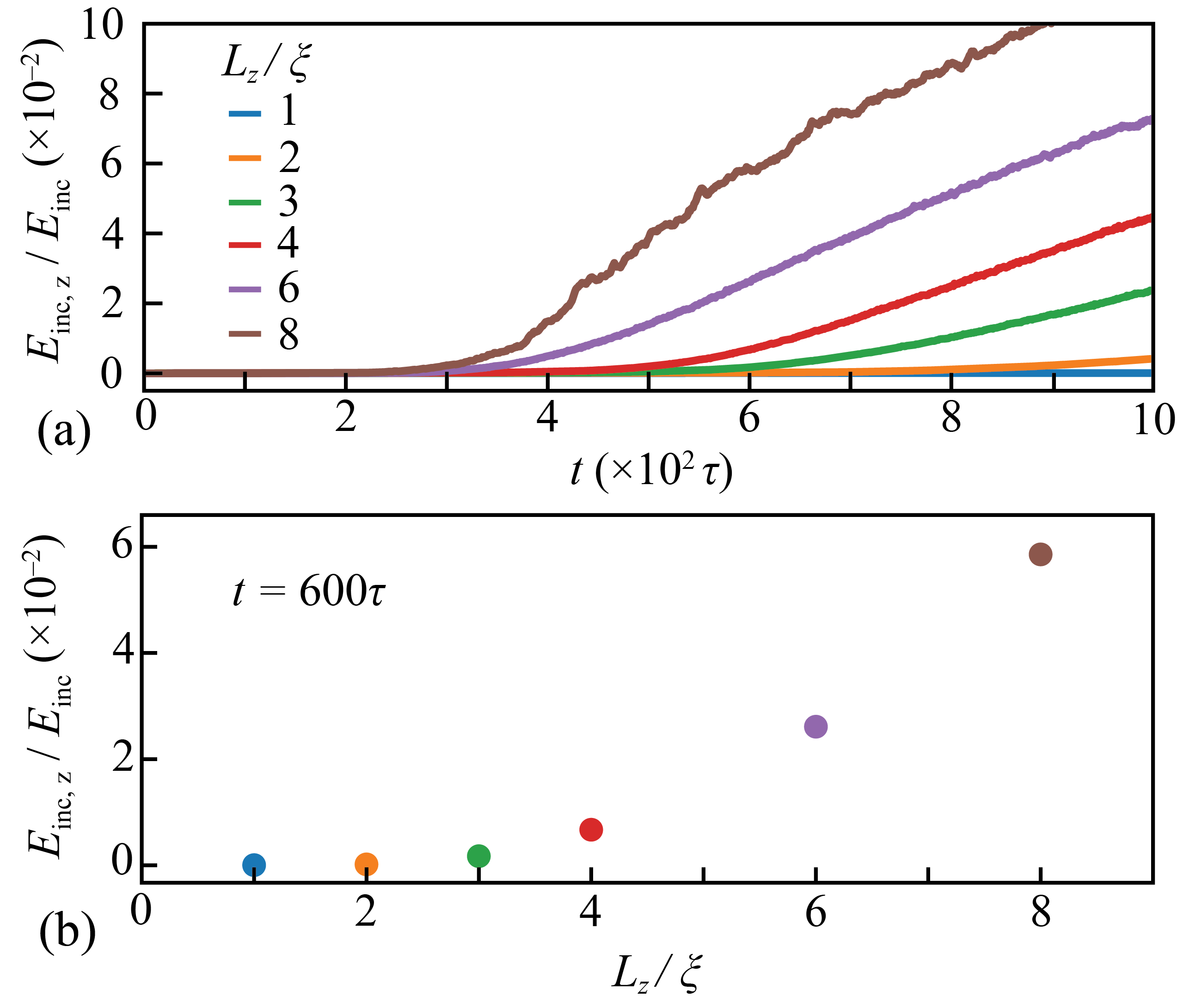}
\caption{ (a) Time evolution of the fraction of the $z$ component
contribution $E_{\mathrm{inc}, z}
/ E_{\mathrm{inc}}$ with various $L_z$ for $N_{\mathrm{vor}} = 20^2$. (b) Dependence of the
fraction $E_{\mathrm{inc}, z} / E_{\mathrm{inc}}$ on $L_z$ at $t = 600 \protect\tau$.}
\label{fig:ratio_ene_inc}
\end{figure}

If the vortices do not curve, the incompressible velocity is confined in the 
$x$-$y$ plane, and $E_{\mathrm{inc},z}$ is negligible. Fig.~\ref%
{fig:ratio_ene_inc} (a) shows the time evolution of the ratio between $E_{%
\mathrm{inc},z}$ and $E_{\mathrm{inc}}$, which can be over $10\%$ with $%
L_{z}=8\xi $ that is a significant contribution. Fig.~\ref{fig:ratio_ene_inc}
(b) shows the $L_{z}$ dependence of the ratio at $t=600\tau $. The
contribution of the z-component is suppressed below $L_{z}=2\xi $ and
rapidly increases above $L_{z}=3\xi $. This result means that the vortex
reconnection effect becomes significant above $L_{z}=3\xi $, agreeing with
the observation in the critical energy observed in Fig. \ref{fig:ene_com} (b).

\paragraph*{Complementary test of vortex-pair annihilation by sound waves.}

To directly demonstrate the energy damping caused by sound waves, the
dynamics of a vortex-pair are analyzed for different cases. In case 1, the
initial BEC has no sound waves, and we implement a vortex-pair with a vortex
distance $d_{\mathrm{V}}(t=0)=3.5\xi $. Here the initial density profile is
given by the two-point Pad\'{e} approximation~\cite{rorai_propagating_2013}.
In case 2, the BEC contains the compressible kinetic energy  $E_{%
\mathrm{com}}\approx 2.5\times 10^{-2}N\mu _{0}$, which is high enough so
that the vortex-pair annihilation may be a three-body process. We then
implement a vortex-pair in the same way as case 1. Fig.~\ref%
{fig:twovortex_evolution} shows the time evolution of the vortex distance $%
d_{\mathrm{V}}(t)$, which is associated with the vortex energy.  $d_{\mathrm{V}(t)}$ is almost constant in case 1, but rapidly varies over a short period and gradually decays over a
long time in case 2, indicating that sound waves may fluctuate the vortex-pair energy
and gradually dissipate it. Note that the energy dissipation due to sound waves does
not need to be strong enough to dampen out the vortex-pair energy by itself
since their role is to destabilize the metastable state. In fact, weak
energy dissipation of thermal excitations' scattering may alter the
vortex-pair annihilation to a two-body process~\cite{baggaley_decay_2018,
groszek_onsager_2016}.

\begin{figure}[t]
\centering
\includegraphics[width=0.43\textwidth]{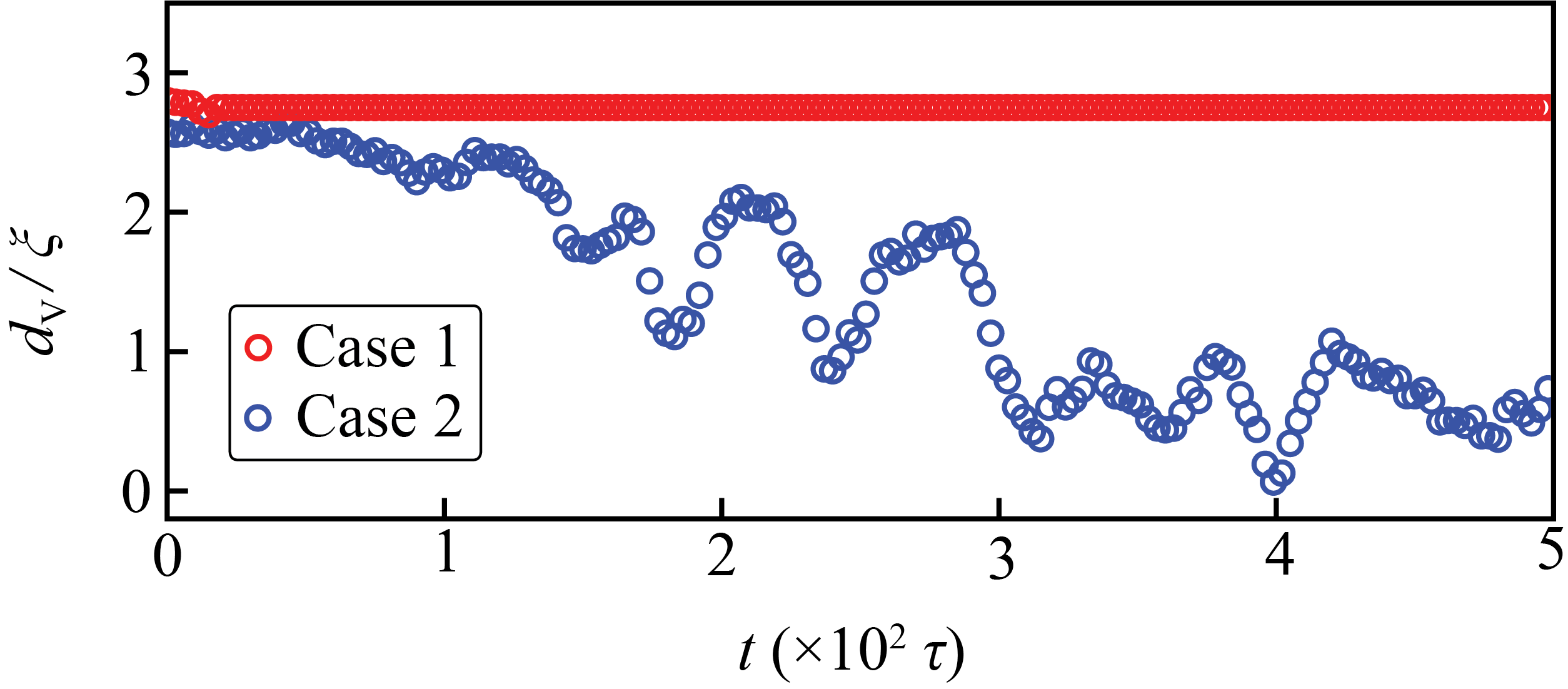}
\caption{Time evolution of the vortex distance $d_V (t)$ in two cases. Case
1 has no initial sound waves, and case 2 has intense sound waves of the
compressible kinetic energy $E_{\mathrm{C}} \approx 2.5 \times 10^{-2} N 
\protect\mu_0$ at $t = 0$. $L_z = 4 \protect\xi$. Here 
$d_V$ is computed for every $t$, and a moving average is taken within
[$t - 10 \protect\tau$, $t + 10 \protect\tau$].}
\label{fig:twovortex_evolution}
\end{figure}

\paragraph{Discussion and conclusion.}

Past studies utilizing the 2D GPE used relatively low initial vortex
densities, therefore the three-body process does not emerge (e.g., $N_{vor}(0) /
\left(L_x L_y\right) = 10^3 / 10^4 \xi^{-2} \approx 153 / 256^2 \xi^{-2}$ in
\cite{baggaley_decay_2018}). On the other hand, the 3D GPE study of the quasi-2D 
BEC \cite{kanai_true_2021} did not observe the
four-body process because of the relatively weak confinement in the third direction, 
leading to a very small critical energy. As a result, the vortex decay dynamics quickly 
pass the transition point from four-body to three-body processes at an early time.

Our simulations focus on the dynamics at the absolute zero temperature, but
rich physics is expected at finite temperatures due to the scattering with
thermal excitations~\cite{baggaley_decay_2018, groszek_onsager_2016}. Quasi-2D BECs have been realized in many experiments \cite{hadzibabic_berezinskiikosterlitzthouless_2006, 2Dexp_2} with the tunable third dimension confinement. The uniform BEC on the 2D plane can be achieved using the box types of potential \cite{boxpot1,boxpot2}. It is also
worth noting that vortex collision can be controlled by localized optical potentials~\cite%
{kwon_sound_2021}, which allow us to measure the sound wave
dissipation to the vortex-pair directly and locally.

In summary, we have demonstrated that the vortex-pair annihilation process
undergoes a dynamical transition from four-body to three-body processes due
to increasing sound waves generated in annihilation
process in the early stage. We characterize the role of the initial vortex density and
the third dimension confinement in the quasi-2D BEC for such
dynamical transition. Our work reveals an elementary mechanism for vortex energy dissipation through pairing annihilation that has been a long-standing issue in the field. 
Understanding such vortex dissipation dynamics allows better characterization of dynamical phenomena (e.g., quantum turbulence) of quantum liquids for new quantum technologies applications.

\begin{acknowledgments}
\emph{Acknowledgement:} 
The authors acknowledge the support of the Air Force Office of Scientific
Research under Grant No. FA9550-20-1-0220 and the National Science
Foundation under Grant No. PHY-2409943, OSI-2228725, ECCS-2411394.
\end{acknowledgments}


\begin{thebibliography}{99}

\bibitem{stock_observation_2005}
S. Stock, Z. Hadzibabic, B. Battelier, M. Cheneau, and J. Dalibard,
Observation of {Phase} {Defects} in {Quasi}-{Two}-{Dimensional} {Bose}--{Einstein} {Condensates},
\href{https://link.aps.org/doi/10.1103/PhysRevLett.95.190403}{Phys. Rev. Lett. \textbf{95}, 190403 (2005).}

\bibitem{hadzibabic_berezinskiikosterlitzthouless_2006}
Z. Hadzibabic, P. Krüger, M. Cheneau, B. Battelier, and J. Dalibard,
Berezinskii--{Kosterlitz}--{Thouless} crossover in a trapped atomic gas,
\href{https://www.nature.com/articles/nature04851}{Nature \textbf{441}, 1118-1121 (2006).}

\bibitem{schweikhard_vortex_2007}
V. Schweikhard, S. Tung, and E. A. Cornell,
Vortex {Proliferation} in the {Berezinskii}-{Kosterlitz}-{Thouless} {Regime} on a {Two}-{Dimensional} {Lattice} of {Bose}-{Einstein} {Condensates},
\href{https://link.aps.org/doi/10.1103/PhysRevLett.99.030401}{Phys. Rev. Lett. \textbf{99}, 030401 (2007).}

\bibitem{kruger_critical_2007}
P. Krüger, Z. Hadzibabic, and J. Dalibard,
Critical {Point} of an {Interacting} {Two}-{Dimensional} {Atomic} {Bose} {Gas},
\href{https://link.aps.org/doi/10.1103/PhysRevLett.99.040402}{Phys. Rev. Lett. \textbf{99}, 040402 (2007).}

\bibitem{nazarenko_bose-einstein_2014}
S. Nazarenko, M. Onorato, and D. Proment,
Bose-{Einstein} condensation and {Berezinskii}-{Kosterlitz}-{Thouless} transition in the two-dimensional nonlinear {Schr\"{o}dinger} model,
\href{https://link.aps.org/doi/10.1103/PhysRevA.90.013624}{Phys. Rev. A \textbf{90}, 013624 (2014).}

\bibitem{sunami_observation_2022}
S. Sunami, V. P. Singh, D. Garrick, A. Beregi, A. J. Barker, K. Luksch, E. Bentine, L. Mathey, and C. J. Foot,
Observation of the {Berezinskii}-{Kosterlitz}-{Thouless} {Transition} in a {Two}-{Dimensional} {Bose} {Gas} via {Matter}-{Wave} {Interferometry},
\href{https://link.aps.org/doi/10.1103/PhysRevLett.128.250402}{Phys. Rev. Lett. \textbf{128}, 250402 (2022).}

\bibitem{scherer_vortex_2007}
D. R. Scherer, C. N. Weiler, T. W. Neely, and B. P. Anderson,
Vortex {Formation} by {Merging} of {Multiple} {Trapped} {Bose}-{Einstein} {Condensates},
\href{https://link.aps.org/doi/10.1103/PhysRevLett.98.110402}{Phys. Rev. Lett. \textbf{98}, 110402 (2007).}

\bibitem{weiler_spontaneous_2008}
C. N. Weiler, T. W. Neely, D. R. Scherer, A. S. Bradley, M. J. Davis, and B. P. Anderson,
Spontaneous vortices in the formation of {Bose}--{Einstein} condensates,
\href{https://www.nature.com/articles/nature07334}{Nature \textbf{455}, 948--951 (2008).}

\bibitem{kanai_flows_2018}
T. Kanai, W. Guo, and M. Tsubota,
Flows with fractional quantum circulation in {Bose}-{Einstein} condensates induced by nontopological phase defects,
\href{https://link.aps.org/doi/10.1103/PhysRevA.97.013612}{Phys. Rev. A \textbf{97}, 013612 (2018).}

\bibitem{kanai_merging_2019}
T. Kanai, W. Guo, and M. Tsubota,
Merging of {Rotating} {Bose}--{Einstein} {Condensates},
\href{http://link.springer.com/10.1007/s10909-018-2110-1}{J. Low Temp. Phys. \textbf{195}, 37--50 (2019).}

\bibitem{ko_kibblezurek_2019}
B. Ko, J. W. Park, and Y. Shin,
Kibble--{Zurek} universality in a strongly interacting {Fermi} superfluid,
\href{https://www.nature.com/articles/s41567-019-0650-1}{Nat. Phys. \textbf{15}, 1227--1231 (2019).}

\bibitem{kanai_torque_2020}
T. Kanai, W. Guo, M. Tsubota, and D. Jin,
Torque and {Angular}-{Momentum} {Transfer} in {Merging} {Rotating} {Bose}-{Einstein} {Condensates},
\href{https://link.aps.org/doi/10.1103/PhysRevLett.124.105302}{Phys. Rev. Lett. \textbf{124}, 105302 (2020).}

\bibitem{goo_defect_2021}
J. Goo, Y. Lim, and Y. Shin,
Defect {Saturation} in a {Rapidly} {Quenched} {Bose} {Gas},
\href{https://link.aps.org/doi/10.1103/PhysRevLett.127.115701}{Phys. Rev. Lett. \textbf{127}, 115701 (2021).}

\bibitem{rysti_suppressing_2021}
J. Rysti, J. T. Mäkinen, S. Autti, T. Kamppinen, G. E. Volovik, and V. B. Eltsov,
Suppressing the {Kibble}-{Zurek} {Mechanism} by a {Symmetry}-{Violating} {Bias},
\href{https://link.aps.org/doi/10.1103/PhysRevLett.127.115702}{Phys. Rev. Lett. \textbf{127}, 115702 (2021).}

\bibitem{barenghi_quantum_2024}
C. F. Barenghi, L. Skrbek, and K. R. Sreenivasan,
\textit{Quantum turbulence},
(Cambridge University Press, Cambridge, 2024).

\bibitem{schole_critical_2012}
J. Schole, B. Nowak, and T. Gasenzer,
Critical dynamics of a two-dimensional superfluid near a nonthermal fixed point,
\href{https://link.aps.org/doi/10.1103/PhysRevA.86.013624}{Phys. Rev. A \textbf{86}, 013624 (2012).}

\bibitem{billam_onsager-kraichnan_2014}
T. P. Billam, M. T. Reeves, B. P. Anderson, and A. S. Bradley,
Onsager-{Kraichnan} {Condensation} in {Decaying} {Two}-{Dimensional} {Quantum} {Turbulence},
\href{https://link.aps.org/doi/10.1103/PhysRevLett.112.145301}{Phys. Rev. Lett. \textbf{112}, 145301 (2014).}

\bibitem{simula_emergence_2014}
T. Simula, M. J. Davis, and K. Helmerson,
Emergence of {Order} from {Turbulence} in an {Isolated} {Planar} {Superfluid},
\href{https://link.aps.org/doi/10.1103/PhysRevLett.113.165302}{Phys. Rev. Lett. \textbf{113}, 165302 (2014).}

\bibitem{lucas_sound-induced_2014}
A. Lucas and P. Surówka,
Sound-induced vortex interactions in a zero-temperature two-dimensional superfluid,
\href{https://link.aps.org/doi/10.1103/PhysRevA.90.053617}{Phys. Rev. A \textbf{90}, 053617 (2014).}

\bibitem{cidrim_controlled_2016}
A. Cidrim, F. E. A. dos Santos, L. Galantucci, V. S. Bagnato, and C. F. Barenghi,
Controlled polarization of two-dimensional quantum turbulence in atomic {Bose}-{Einstein} condensates,
\href{https://link.aps.org/doi/10.1103/PhysRevA.93.033651}{Phys. Rev. A \textbf{93}, 033651 (2016).}

\bibitem{groszek_onsager_2016}
A. J. Groszek, T. P. Simula, D. M. Paganin, and K. Helmerson,
Onsager vortex formation in {Bose}--{Einstein} condensates in two-dimensional power-law traps,
\href{https://link.aps.org/doi/10.1103/PhysRevA.93.043614}{Phys. Rev. A \textbf{93}, 043614 (2016).}

\bibitem{yu_theory_2016}
X. Yu, T. P. Billam, J. Nian, M. T. Reeves, and A. S. Bradley,
Theory of the vortex-clustering transition in a confined two-dimensional quantum fluid,
\href{https://link.aps.org/doi/10.1103/PhysRevA.94.023602}{Phys. Rev. A \textbf{94}, 023602 (2016).}

\bibitem{seo_observation_2017}
S. W. Seo, B. Ko, J. H. Kim, and Y. Shin,
Observation of vortex-antivortex pairing in decaying {2D} turbulence of a superfluid gas,
\href{https://www.nature.com/articles/s41598-017-04122-9}{Sci. Rep. \textbf{7}, 4587 (2017).}

\bibitem{karl_strongly_2017}
M. Karl and T. Gasenzer,
Strongly anomalous non-thermal fixed point in a quenched two-dimensional {Bose} gas,
\href{https://iopscience.iop.org/article/10.1088/1367-2630/aa7eeb}{New J. Phys. \textbf{19}, 093014 (2017).}

\bibitem{groszek_vortex_2018}
A. J. Groszek, M. J. Davis, D. M. Paganin, K. Helmerson, and T. P. Simula,
Vortex {Thermometry} for {Turbulent} {Two}-{Dimensional} {Fluids},
\href{https://link.aps.org/doi/10.1103/PhysRevLett.120.034504}{Phys. Rev. Lett. \textbf{120}, 034504 (2018).}

\bibitem{frishman_turbulence_2018}
A. Frishman and C. Herbert,
Turbulence {Statistics} in a {Two}-{Dimensional} {Vortex} {Condensate},
\href{https://link.aps.org/doi/10.1103/PhysRevLett.120.204505}{Phys. Rev. Lett. \textbf{120}, 204505 (2018).}

\bibitem{gauthier_giant_2019}
G. Gauthier, M. T. Reeves, X. Yu, A. S. Bradley, M. A. Baker, T. A. Bell, H. Rubinsztein-Dunlop, M. J. Davis, and T. W. Neely,
Giant vortex clusters in a two-dimensional quantum fluid,
\href{https://www.science.org/doi/10.1126/science.aat5718}{Science \textbf{364}, 1264-1267 (2019).}

\bibitem{johnstone_evolution_2019}
S. P. Johnstone, A. J. Groszek, P. T. Starkey, C. J. Billington, T. P. Simula, and K. Helmerson,
Evolution of large-scale flow from turbulence in a two-dimensional superfluid,
\href{https://www.science.org/doi/10.1126/science.aat5793}{Science \textbf{364}, 1267-1271 (2019).}

\bibitem{sachkou_coherent_2019}
Y. P. Sachkou, C. G. Baker, G. I. Harris, O. R. Stockdale, S. Forstner, M. T. Reeves, X. He, D. L. McAuslan, A. S. Bradley, M. J. Davis, and W. P. Bowen,
Coherent vortex dynamics in a strongly interacting superfluid on a silicon chip,
\href{https://www.science.org/doi/10.1126/science.aaw9229}{Science \textbf{366}, 1480-1485 (2019).}

\bibitem{groszek_decaying_2020}
A. J. Groszek, M. J. Davis, and T. P. Simula,
Decaying quantum turbulence in a two-dimensional {Bose}-{Einstein} condensate at finite temperature,
\href{https://scipost.org/10.21468/SciPostPhys.8.3.039}{Sci{Post} Phys. \textbf{8}, 039 (2020).}

\bibitem{reeves_turbulent_2022}
M. T. Reeves, K. Goddard-Lee, G. Gauthier, O. R. Stockdale, H. Salman, T. Edmonds, X. Yu, A. S. Bradley, M. Baker, H. Rubinsztein-Dunlop, M. J. Davis, and T. W. Neely,
Turbulent {Relaxation} to {Equilibrium} in a {Two}-{Dimensional} {Quantum} {Vortex} {Gas},
\href{https://link.aps.org/doi/10.1103/PhysRevX.12.011031}{Phys. Rev. X \textbf{12}, 011031 (2022).}

\bibitem{easton_vortex_2023}
T. Easton, M. Kokmotos, and G. Barontini,
Vortex clustering in trapped {Bose}--{Einstein} condensates,
\href{https://www.nature.com/articles/s41598-023-46549-3}{Sci. Rep. \textbf{13}, 19432 (2023).}

\bibitem{kwon_relaxation_2014}
W. J. Kwon, G. Moon, J. Choi, S. W. Seo, and Y. Shin,
Relaxation of superfluid turbulence in highly oblate {Bose}-{Einstein} condensates,
\href{https://link.aps.org/doi/10.1103/PhysRevA.90.063627}{Phys. Rev. A \textbf{90}, 063627 (2014).}

\bibitem{stagg_generation_2015}
G. W. Stagg, A. J. Allen, N. G. Parker, and C. F. Barenghi,
Generation and decay of two-dimensional quantum turbulence in a trapped {Bose}-{Einstein} condensate,
\href{https://link.aps.org/doi/10.1103/PhysRevA.91.013612}{Phys. Rev. A \textbf{91}, 013612 (2015).}

\bibitem{baggaley_decay_2018}
A. W. Baggaley and C. F. Barenghi,
Decay of homogeneous two-dimensional quantum turbulence,
\href{https://link.aps.org/doi/10.1103/PhysRevA.97.033601}{Phys. Rev. A \textbf{97}, 033601 (2018).}

\bibitem{nazarenko_freely_2007}
S. Nazarenko and M. Onorato,
Freely decaying {Turbulence} and {Bose}--{Einstein} {Condensation} in {Gross}--{Pitaevski} {Model},
\href{http://link.springer.com/10.1007/s10909-006-9271-z}{J. Low Temp. Phys. \textbf{146}, 31-46 (2007).}

\bibitem{smirnov_dynamics_2012}
L. A. Smirnov and V. A. Mironov,
Dynamics of two-dimensional dark quasisolitons in a smoothly inhomogeneous {Bose}-{Einstein} condensate,
\href{https://link.aps.org/doi/10.1103/PhysRevA.85.053620}{Phys. Rev. A \textbf{85}, 053620 (2012).}

\bibitem{smirnov_scattering_2015}
L. A. Smirnov and A. I. Smirnov,
Scattering of two-dimensional dark solitons by a single quantum vortex in a {Bose}-{Einstein} condensate,
\href{https://link.aps.org/doi/10.1103/PhysRevA.92.013636}{Phys. Rev. A \textbf{92}, 013636 (2015).}

\bibitem{rickinson_diffusion_2018}
E. Rickinson, N. G. Parker, A. W. Baggaley, and C. F. Barenghi,
Diffusion of quantum vortices,
\href{https://link.aps.org/doi/10.1103/PhysRevA.98.023608}{Phys. Rev. A \textbf{98}, 023608 (2018).}

\bibitem{kanai_true_2021}
T. Kanai and W. Guo,
True {Mechanism} of {Spontaneous} {Order} from {Turbulence} in {Two}-{Dimensional} {Superfluid} {Manifolds},
\href{https://link.aps.org/doi/10.1103/PhysRevLett.127.095301}{Phys. Rev. Lett. \textbf{127}, 095301 (2021).}

\bibitem{ma_phase_2019}
T. Ma and S. Wang,
\textit{Phase {Transition} {Dynamics}},
(Springer International Publishing, Cham, 2019).

\bibitem{pethick_bose-einstein_2008}
C. J. Pethick and H. Smith,
\textit{Bose-{Einstein} condensation in dilute gases},
2nd ed.,
(Cambridge University Press, Cambridge, 2008).

\bibitem{rooney_suppression_2011}
S. J. Rooney, P. B. Blakie, B. P. Anderson, and A. S. Bradley,
Suppression of {Kelvon}-induced decay of quantized vortices in oblate {Bose}-{Einstein} condensates,
\href{https://link.aps.org/doi/10.1103/PhysRevA.84.023637}{Phys. Rev. A \textbf{84}, 023637 (2011).}

\bibitem{chomaz_emergence_2015}
L. Chomaz, L. Corman, T. Bienaimé, R. Desbuquois, C. Weitenberg, S. Nascimbène, J. Beugnon, and J. Dalibard,
Emergence of coherence via transverse condensation in a uniform quasi-two-dimensional {Bose} gas,
\href{https://www.nature.com/articles/ncomms7162}{Nat Commun \textbf{6}, 6162 (2015).}

\bibitem{abramowitz_handbook_2013}
M. Abramowitz and I. A. Stegun (Eds.),
\textit{Handbook of mathematical functions: with formulas, graphs, and mathematical tables},
(Dover Publ, New York, NY, 2013).

\bibitem{banerjee_structural_2022}
R. Boral, S. Sarkar, and P. K. Mishra,
Structural {Transformation} and {Melting} of the {Vortex} {Lattice} in the {Rotating} {Bose} {Einstein} {Condensates}. In: Banerjee, S., Saha, A. (eds) Nonlinear Dynamics and Applications. Springer Proceedings in Complexity. Springer, Cham. 
\href{https://link.springer.com/10.1007/978-3-030-99792-2_106}{https://link.springer.com/10.1007/978-3-030-99792-2\_106}

\bibitem{numasato_direct_2010}
R. Numasato, M. Tsubota, and V. S. L’vov,
Direct energy cascade in two-dimensional compressible quantum turbulence,
\href{https://link.aps.org/doi/10.1103/PhysRevA.81.063630}{Phys. Rev. A \textbf{81}, 063630 (2010).}

\bibitem{bradley_energy_2012}
A. S. Bradley and B. P. Anderson,
Energy {Spectra} of {Vortex} {Distributions} in {Two}-{Dimensional} {Quantum} {Turbulence},
\href{https://link.aps.org/doi/10.1103/PhysRevX.2.041001}{Phys. Rev. X \textbf{2}, 041001 (2012).}

\bibitem{horng_two-dimensional_2009}
T.-L. Horng, C.-H. Hsueh, S.-W. Su, Y.-M. Kao, and S.-C. Gou,
Two-dimensional quantum turbulence in a nonuniform {Bose}--{Einstein} condensate,
\href{https://link.aps.org/doi/10.1103/PhysRevA.80.023618}{Phys. Rev. A \textbf{80}, 023618 (2009).}


\bibitem{rorai_propagating_2013}
C. Rorai, K. R. Sreenivasan, and M. E. Fisher,
Propagating and annihilating vortex dipoles in the {Gross}-{Pitaevskii} equation,
\href{https://link.aps.org/doi/10.1103/PhysRevB.88.134522}{Phys. Rev. B \textbf{88}, 134522 (2013).}


\bibitem{2Dexp_2} C.-L. Hung, X. Zhang, N. Gemelke, and C. Chin, Observation of scale invariance and universality in two-dimensional Bose gases, \href{https://www.nature.com/articles/nature09722}{Nature \textbf{470}, 236 (2011). }

\bibitem{boxpot1} A. L. Gaunt, T. F. Schmidutz, I. Gotlibovych, R. P. Smith, Z. Hadzibabic, Bose-Einstein condensation of atoms in a uniform potential, \href{
https://doi.org/10.1103/PhysRevLett.110.200406}{Phys. Rev. Lett. \textbf{110}, 200406 (2013). }

\bibitem{boxpot2} B. Mukherjee, Z. Yan, P. B. Patel, Z. Hadzibabic, T. Yefsah, J. Struck, and M. W. Zwierlein, Homogeneous Atomic Fermi Gases, \href{https://doi.org/10.1103/PhysRevLett.118.123401}{Phys. Rev. Lett. \textbf{118}, 123401 (2017). }

\bibitem{kwon_sound_2021}
W. J. Kwon, G. Del Pace, K. Xhani, L. Galantucci, A. Muzi Falconi, M. Inguscio, F. Scazza, and G. Roati,
Sound emission and annihilations in a programmable quantum vortex collider,
\href{https://www.nature.com/articles/s41586-021-04047-4}{Nature \textbf{600}, 64 (2021).}

\end{thebibliography}
\end{document}